\documentstyle[12pt]{article}
\newcommand{\bce}{\begin{center}}
\newcommand{\ece}{\end{center}}
\newcommand{\beq}{\begin{equation}}
\newcommand{\eeq}{\end{equation}}
\newcommand{\bea}{\vspace{0.25cm}\begin{eqnarray}}
\newcommand{\eea}{\end{eqnarray}}

\newcommand{\ba}{\begin{array}}
\newcommand{\ea}{\end{array}}

\newcommand{\doublespace}{
    \renewcommand{\baselinestretch}{1.6}\large\normalsize}

\def\lsim{\mathrel{\rlap{\lower4pt\hbox{\hskip1pt$\sim$}}
    \raise1pt\hbox{$<$}}}	  %less than or approx. symbol
\def\gsim{\mathrel{\rlap{\lower4pt\hbox{\hskip1pt$\sim$}}
    \raise1pt\hbox{$>$}}}	  %greater than or approx. symbol

\def\Pom{{\bf I\!P}}

\def\lsim{\mathrel{\rlap{\lower4pt\hbox{\hskip1pt$\sim$}}
    \raise1pt\hbox{$<$}}}         %less than or approx. symbol
\def\gsim{\mathrel{\rlap{\lower4pt\hbox{\hskip1pt$\sim$}}
    \raise1pt\hbox{$>$}}}         %greater than or approx. symbol

\def\Pom{{\bf I\!P}}

  \advance\oddsidemargin  by -1in
  \advance\evensidemargin by -1in
  \advance\topmargin      by -0.5in
\def\lsim{\mathrel{\rlap{\lower4pt\hbox{\hskip1pt$\sim$}}
    \raise1pt\hbox{$<$}}}         %less than or approx. symbol
\def\gsim{\mathrel{\rlap{\lower4pt\hbox{\hskip1pt$\sim$}}
    \raise1pt\hbox{$>$}}}         %greater than or approx. symbol

\def\Pom{{\bf I\!P}}
\def\beq{\begin{equation}}
\def\endeq{\end{equation}}
\def\arr{\begin{eqnarray}}
\def\endarr{\end{eqnarray}}
\makeindex

\begin{document}

\phantom.\hspace{8.8cm}{\Large\bf KFA-IKP(TH)-1994-13\bigskip\\}
\phantom.\hspace{10.2cm}{\Large \bf 16 May   1994}\vspace{1.5cm}\\
\begin{center}
{\bf \huge
Why more hadronlike photons \\
produce less particles
on nuclear targets \vspace{1.0cm}}

{\Large N.N.Nikolaev$^{a,b)}$, B.G.Zakharov$^{b)}$ and
V.R.Zoller$^{c)}$\medskip\\ }
{\large \it
$^{a)}$Institut  f\"ur Kernphysik, Forschungszentrum J\"ulich,\\
D-52425 J\"ulich, Germany\medskip\\
$^{b)}$L.D.Landau Institute for Theoretical Physics,\\
GSP-1, 117940, ul. Kosygina 2, 117334 Moscow, Russia\medskip\\
$^{c)}$Institute for Theoretical and Experimental Physics,\\
ul. B.Chermushkinskaya 25, 117259 Moscow, Russia\vspace{1cm}\\
}
{\bf           Abstract}
\end{center}
In hadron-nucleus interactions, the stronger is nuclear shadowing
in the total cross section the higher is the multiplicity of
secondary hadrons. In deep inelastic scattering, nuclear
shadowing at small $x$ is associated with the hadronlike behaviour
of photons as contrasted to the pointlike behaviour in the
non-shadowing region of large $x$. In this paper we
predict smaller
mean multiplicity of secondary hadrons, and weaker fragmentation
of the target nucleus, in deep inelastic leptoptoproduction
on nuclei in the shadowing region of small $x$ as compared
to the non-shadowing region of large $x$. This paradoxial
conclusion has its origin in nuclear enhancement of the coherent
diffraction
dissociation of photons. We present numerical predictions for
multiproduction in $\mu Xe$ interactions studied by the Fermilab
E665 collaboration.
\pagebreak\\

\doublespace
%-----------------------------------------------

%                             Section 1

%---------------------

\section{Introduction}

At high energies $\nu$ and/or very large ${1\over x}\gg 1$, where
$x$ is the Bjorken variable,
the real, and virtual, photoabsorption can conveniently be considered
as interaction with the target nucleon (nucleus) of
hadronic (multiparton) Fock
states $X$ of the photon (for the review and references to the early
work see [1]). The best known
consequence of this hadronic Fock-state mediated photoabsorption
is that nuclear shadowing in the forward Compton scattering
off nuceli, i.e., in the nuclear structure function, will be
similar to that in the hadronic $XA$ scattering [2]. In the
framework of multiple scattering theory [3], nuclear
shadowing has its origin in diffractive excitation $\gamma^{*}
\rightarrow X$ of the intermediate state $X$ followed by its
deexcitation $X\rightarrow \gamma^{*}$ inside the nucleus [4].
The closely related process is the direct diffractive excitation
of hadronic Fock states of the photon $\gamma^{*}+N(A) \rightarrow
N(A)+X$, which is inseparable from nuclear shadowing. Although
at large
$Q^{2}$  deep inelastic scattering (DIS) will eventually be dominated
by interactions of the small-size  multiparton Fock states, the
remarkable feature of QCD is that hadronic-size Fock components
of the photon contribute significantly even at asymptotically
large $Q^{2}$, and completely dominate diffraction dissociation
of photons [5,6]. In this paper we discuss a novel, and paradoxial,
feature of diffractive DIS on nuclei: weaker multiproduction
of secondary hadrons, and weaker fragmentation of the target
nucleus, in the shadowing region of $x\ll 1$, when
photons become hadronlike, with respect to the multiproduction
in the nonshadowing region of large $x$, where photons are
pointlike.

Because of nuclear shadowing, the nuclear cross section is
smaller than $A$ times the free-nucleon cross section (hereafter
$A$ is the nuclear mass number). In the multiple scattering theory
[3], this shadowing in the hadron-nucleus interaction
comes from multiple intranuclear rescatterings
of the projectile. A convenient parameter which measures the
strength of multiple rescatterings is (for the review see [1])
\beq
\bar{\nu}={A\sigma_{N}\over \sigma_{A}}\, .
\label{eq:1.1}
\endeq
Multiple intranuclear interactions are one of the sources of stronger
multiproduction on nuclei, and a good empirical approximation for the
scaled nuclear multiplicity of secondary particles
$R=\langle n_{A} \rangle / \langle n_{N}
\rangle $ is
\beq
R\approx {1\over 2}(1+\bar{\nu})\, .
\label{eq:1.2}
\endeq
The larger is the free-nucleon cross section $\sigma_{hN}$ the
stronger is the nuclear shadowing (for the review see [1,7]).
For instance, because $\sigma_{pN}>\sigma_{\pi N}>\sigma_{KN}$,
one has $\bar{\nu}_{pA}>\bar{\nu}_{\pi A}>\bar{\nu}_{KA}$, and
indeed experimentally the scaled nuclear multiplicity satisfies
$R_{pA}  > R_{\pi A} > R_{K A} $ (for the review see [1]).

For the weakly interacting pointlike probes, like in DIS
in the non-shadowing (NS) region, the impulse
approximation is exact and the amplitude of forward elastic
scattering on a nucleus equals A times the free-nucleon amplitude:
\beq
F_{A}=AF_{N}\,.
\label{eq:1.3}
\endeq
This implies $\bar{\nu}=1$. Furthermore, because by virtue of the
optical theorem, the $n$-particle production cross section is
related to the $n$-particle discontinutity of the forward scattering
amplitude. Then,
if taken at face value,
Eq.~(\ref{eq:1.3}) would have
imlpied identical discontinuities of the nuclear and the free-nucleon
amplitudes, i.e., identical multiplicities for the nuclear and
free-nucleon interactions. This naive expectation fails, though,
because of the cascading effects [1,2,8-12], which do not affect
the nuclear cross section, but contribute to the particle production.
The significance of cascading as a necessary condition for
thermalization of the produced particles and for the formation of
the quark-gluon plasma in collisions of ultrarelativistic heavy
ions is discussed in [12]. There is a mounting experimental
evidence for cascading in leptoproduction on nuclei [12,13].

It was suggested quite a time ago [1,2,8,9], that in the
leptoproduction on nuclei, one can control $\bar{\nu}$ by varying
the Bjorken variable $x$ from the non-shadowing (NS) region of
large $x \gsim 0.05$, where $\bar{\nu}=1$, to the shadowing
(SH) region of very small $x$, where $\bar{\nu}>1$. For instance,
in the $\mu Xe$ scattering at $x\sim 10^{-3}$ the shadowing effect
is rather strong, $\bar{\nu}\sim 1.5$ [14-16]. Then, the empirical
law Eq.~(\ref{eq:1.2}) would have suggested a strong, $\sim 25\%$,
nuclear enhancement of the mean multiplicity. Apart from the larger
mean multiplicity of secondary paricles, in the shadowing region
of small $x$ one would naively have expected other signals of
enhanced intranuclear reinteractions like the higher multiplicity
of knocked-out protons (grey tracks) and of the nucleus
fragmentation in general, which rise with $\bar{\nu}$ [1,2,8,9].

In this communication we wish to demonstrate how
the above summarized conventional wisdom fails:
 the very mechanism
of the hadronlike behavior of (virtual) photons leads to a weaker
multiproduction on, and weaker fragmentation of,
nuclei in the shadowing region as compared
to multiproduction in the non-shadowing region. The principal
observation goes as follows: In the free-nucleon interactions,
the diffraction dissociation (DD) events a characterized by a
large (pseudo)rapidity gap (LRG) between the recoil proton and
the hadronic debris from the diffraction dissociation of
photons. Because of this large rapidity gap,
the DD events have smaller mean
multiplicity than the non-diffraction dissociation (ND) events.
The major finding of the present paper is that, in
DIS  on nuclei, the fraction of DD and/or
LRG events significantly rises with $A$. On the
black-disc nucleus,
the coherent DD, which leaves the target nucleus in the ground
state and consequently gives a vanishing hadronic activity
in the nucleus fragmentation region,
will make $\sim 50\%$ of the total DIS
cross section. Nuclear shadowing and  DD come in one
package, and because of nuclear enhancement of DD the
hadronlike photons produce less secondary hadrons than the
pointlike photons. Recently,
there was much theoretical interest in  DD of photons [5,6], and LRG
events were observed in DIS at HERA [17]
with the rate which agrees with the theoretical prediction [5].
The novel manifestation of DD of photons, discussed in this paper,
adds to the growing interest in the large-rapidity gap physics
in DIS.

The observation of different $A$-dependence of the diffractive
and nondiffractive multiproduction in the hadron-nuclei collisions
was made by one of the authors quite a time ago [18]. The major
difference between the leptoproduction and hadroproduction is
that in the latter case the nuclear DD  only makes
a small fraction of the nuclear cross section, whereas in
the leptoproduction DD cross section is much larger.

The paper is organized as follows. In section 2 we start with
the brief review of the dipole-cross section approach to
diffractive DIS. In section 3 we derive
the cross section for the coherent and incoherent DD
on nuclei and discuss the relationship between
the DD of photons and nuclear shadowing.
We also demonstrate the nuclear enhancement of
DD cross section. In section 4 we
discuss the impact of finite energy effects on the so-called
triple-pomeron component of the coherent and incoherent
DD on nuclei and on nuclear shadowing,
and present predictions for the $A$ dependence of the rapidity
gap distribution.
The experimental signatures
of nuclear enhancement of diffraction dissociation are discussed
in section 5.
In section 6 we comment on why the effects of DD in the
leptoproduction and hadroproduction on nuclei are so much different.
Our principal
results and conlcusions are summarized in section 7.

%----------------------------------------------

%                                  Section 2

%-----------------

\section{Hadronic prioperties of the photon and the dipole-cross
section representation}

We start with the brief review of the dipole-cross section
representation for diffractive DDIS
[14,5,6],
which provides a unified description of nuclear shadowing and of
diffraction dissociation of photons. At small $x$, DIS
can be viewed as interaction of the hadronic fluctuations
the virtual photon transforms into at large distance
\beq
\Delta z \sim {2\nu \over Q^{2}+M^{2}} \sim {1 \over m_{p}x} \gsim
R_{N},R_{A}
\label{eq:2.1}
\endeq
in front of the target nucleon (nucleus) [1]. Here $\nu$
and $Q^{2}$
are the laboratrory energy and virtuality of the photon,
 $M$ is the invariant
mass of hadronic fluctuation of the photon and $R_{A,N}$ is the
radius of the target nucleus (nucleon). Because of $\Delta z
\gsim R_{A}$, the transverse separation $\vec{r}$
of partons in the multiparton
Fock state of the photon becomes as good a conserved quantity as
the angular momentum. The resulting diagonalization of the
diffractive $S$-matrix in the $\vec{r}$-representation leads to
a very simple, and intuitively appealing, description of
diffractive interactions in the
dipole-cross section approach.

We present the approach
starting with interactions of the simplest $q\bar{q}$ Fock state of
the photon. The principal quantities are the total cross section
$\sigma(r)$ for interaction of the colour dipole, i.e., the
colour-singlet $q\bar{q}$ pair with the transvrese separation
$\vec{r}$ with the nucleon target, and the wave functions
$| \Psi_{T,L}(\alpha,\vec{r}\,)|^{2}$ for the (T) transverse and
(L) longitudinal photons, computed in [14]. Here $\alpha$ is a
fraction of the lightcone momentum of the photon carried by
the quark of the $q\bar{q}$ pair.
The total photoabsorption
cross section and the inclusive forward DD cross section for the
free-nucleon target are given by
\beq
\sigma_{T,L} = \int_{0}^{1}d\alpha \int d^{2}\vec{r}
\>|\Psi_{T,L}(\alpha,\vec{r}\,)| ^{2}
\sigma (r) =\left< \sigma(r) \right>_{T,L}
\label{eq:2.2}
\endeq
\beq
\left. {d\sigma_{D}^{(N)} \over
dt }\right|_{t=0}=
\int dM^{2}\left. {d\sigma_{D}^{(N)} \over
dM^{2}dt }\right|_{t=0}
 = \int_{0}^{1}d\alpha \int d^{2}\vec{r}\>
\vert \Psi_{T,L}(\alpha,\vec{r}\,)\vert ^{2}
{ {\sigma (r)}^{2} \over 16\pi } =
{1\over 16\pi} \left< \sigma(r)^{2}\right>_{T,L}
\label{eq:2.3}
\endeq

In the diffraction production of the
state of mass $M$ the target proton receives a small recoil momentum
$\kappa$  which in the laboratory frame equals
\beq
\kappa = {M^{2}+Q^{2} \over 2m_{N}\nu}=m_{N}x(1+{M^{2}\over Q^{2}})=
m_{N}x[1+\exp(y)]\, ,
\label{eq:2.4}
\endeq
Here
\beq
y=\log\left({M^{2}\over Q^{2}}\right)
\label{eq:2.5}
\endeq
is a convenient variable which measures the mass of the diffractively
excited state.  As we shall see below, DD has simple scaling
properties in terms of this variable $y$, and we strongly advocate
an analysis of DD in terms of this new variable.
Let $W$ be the total collision energy in the photon-proton c.m.s,
$W^{2}=2m_{p}\nu - Q^{2}$. The recoil proton
emerges in the final state separated from the hadronic debris
of the photon by large (pseudo)rapidity gap
\beq
\Delta \eta \approx \log\left({W^{2} \over M^{2}}\right)=
\log({1\over x})-\log({M^{2}\over Q^{2}}) =\log({1\over x})-y
\label{eq:2.6}
\endeq
For the reaction to be the diffraction dissociation, the
(pseudo)rapidity gap $\Delta\eta$ must be large, $\Delta\eta
\gsim 2.5-3$ (for the recent review on diffraction dissociation
in hadronic scattering see [19]).
 The total (pseudo)rapidity span equals
\beq
Y_{max}\approx \log({1\over x})+
\log({Q^{2}\over \langle p_{\perp}\rangle^{2}})\, ,
\label{eq:2.7.}
\endeq
where $\langle p_{\perp} \rangle$ is the mean transvrese momentum
of secondary hadrons, $\langle p_{\perp} \rangle
\sim {1\over 2}m_{\rho}$.
The maximal kinematically allowed rapidity gap $\Delta\eta\sim
Y_{max}$ corresponds to exclusive production of the very low-mass
state like the continuum two-pion state near the threshold and/or
the $\rho^{0}$ meson, $M_{min}\sim {1\over 2}m_{\rho}$, and in DIS
\beq
y_{min}=\log\left({M_{min}^{2}\over Q^{2}}\right) < 0\,.
\label{eq:2.8}
\endeq
Notice, that $y=0$ corresponds to the rapidity of the virtual photon.

Hereafter we concentrate on the dominant DD
of transverese photons and suppress the subscript T.
Excitation of $q\bar{q}$ pairs leads to the mass spectrum peaked
at $M^{2} \sim Q^{2}$ [5]:
\beq
\left.{d\sigma_{D}^{(N)} \over dtdM^{2} }\right|_{t=0}
 \approx
    \Sigma_{DD} { M^{2} \over (Q^{2}+M^{2})^{3} }\, .
\label{eq:2.9}
\endeq
DD of photons can also be viewed as DIS on pomerons ($\Pom$),
and the mass
spectrum can be related to the pomeron structure function.
Diffraction excitation of the $q\bar{q}$ Fock state of the photon
corresponds to DIS on the valence $q\bar{q}$ component of the
pomeron, with the structure function [5,6] $F_{2}^{\Pom}(\beta)
\propto \beta(1-\beta)$, where $\beta = Q^{2}/(Q^{2}+M^{2})$ is
the Bjorken variable for the $e\Pom$ deep inelastic scattering.
This is the smooth spectrum, which does not contain an explicit
$\rho^{0}$ resonance contribution, but it correctly reproduces
the resonance-smeared mass spectrum even in the limit of real
photoproduction $Q^{2}=0$ [5,20]. The normalization of the
mass spectrum $\; \Sigma_{DD}\;$ is
such that the integrated diffraction excitation cross section
\beq
\sigma_{D}^{(N)} = r_{D}\sigma_{N}=\int dM^{2} dt
\left. {d\sigma_{D} \over dtdM^{2} }\right|_{t=0}
\approx
\int dM^{2}
{1\over b_{D}} \left.
 {d\sigma_{D} \over dtdM^{2} }\right|_{t=0}
\label{eq:2.10}
\endeq
makes the fraction $r_{D}\sim 10\%$ of the total photoabsorption
on free nucleons. This fraction $r_{D}$ was predicted to only
weakly depend on $x$ and $Q^{2}$ [5,21], which is in good
agreement with the first experimental data on DD of photons
from the ZEUS collaboration at HERA [17]. The above estimate
$r_{D}\sim 10\%$ for excitation of $q\bar{q}$ Fock states of
the photon is found [5,21] with the dipole cross section
$\sigma(r)$ of Ref.~[14] and for the diffraction slope
$b_{D} \sim b_{\pi N} \sim 10$GeV$^{-2}$. The effects
of higher $q\bar{q}g,...$ Fock states will be discussed below.

Let $n(W^{2})$ be the mean multiplicity in the generic inelastic
interaction. Then, in the DD events the mean multiplicity will
be approximately equal to
\beq
n_{D}(M^{2})\approx
n(W^{2}\exp(-\Delta \eta)) < n(W^{2})\, .
\label{eq:2.11}
\endeq
Consequently, the enhancement of the fraction of DD
events in DIS on nuclei results in the lower mean multiplicity
of secondary particles in the diffractive
multiproduction on nuclei. In the
next section we shall demonstrate that such an enhancement indeed
takes place. Another important signature of DD is a very small
recoil of the target nucleon (nucleus), which means a lack of
any observable hadronic activity in the target region.

%---------------------------------------------------

%                                Section 3

%--------------------

\section{Hadronic properties of the photon and deep inelastic
scattering off nuclei}

In this section we discuss the impact of diffraction dissociation
in the leptoproduction off nuclei, starting with the high-energy
limit of $\Delta z \gg R_{A}$. In the
interaction with nuclear targets one has to distinguish
the (coh) coherent diffraction dissociation $\gamma^{*}+A
\rightarrow X+A$, when the target nucleus remains in the ground
state, and the (inc) incoherent diffraction dissociation
$\gamma^{*}+A \rightarrow X+A^{*}$, when one summs over all
excitations and breakup of the traget nucleus not followed by
the secondary
particle production in the nucleus fragmentation region.
The both processes lead to the LRG events, but have
different $A$-dependence and slightly
different dependence on the
rapidity gap $\Delta \eta$.

We derive the cross sections for the coherent and incoherent
diffraction dissociation on a nucleus using the technique
developed in [22,23]. We start with the total cross section
of photoabsorption on a nucleus which equals
\arr
\sigma_{A}=R_{sh}A\sigma_{N}=A\sigma_{N}-\Delta\sigma_{sh}=
2\left< \int d^{2}\vec{b}\{
1-\exp[-{1\over 2}\sigma(r)T(b)]\}
\right>\nonumber\\   =
A\langle \sigma(r)\rangle -
\int d^{2}\vec{b}\left<
\exp[-{1\over 2}\sigma(r)T(b)]
\{1-[1-
{1\over 2}\sigma(r)T(b)]
\exp[{1\over 2}\sigma(r)T(b)]\}\right>\nonumber\\
=A\left< \sigma(r)\right> -
{1\over 4}\int d^{2}\vec{b}T(b)^{2}\left<
\sigma(r)^{2}\exp[-{1\over 2}\sigma(r)T(b)]
\right> + ...
\, .
\label{eq:3.1}
\endarr
Here $\vec{b}$ is the impact parameter, $T(b)=\int dz
n_{A}(z,\vec{b})$ is the optical thickness of the nucleus
and $n_{A}(z,\vec{b})$ is the nuclear matter density (for
the nuclear density parametrizations see [24]). In
Eq.~(\ref{eq:3.1}) we decomposed the nuclear cross section
into the impulse approximation term $A\sigma_{N}$ and the
shadowing term $\Delta\sigma_{sh}$.
Although the cross section of photoabsorption
on the free nucleon is small,
\beq
\sigma_{N}={4\pi^{2}\alpha_{em} \over Q^{2}}F_{2}(x,Q^{2})\, ,
\label{eq:3.2}
\endeq
and rapidly vanishes with rising $Q^{2}$, the nuclear shadowing
persists at all $Q^{2}$ and the ratio $\Delta\sigma_{sh}/\sigma_{A}=
1-R_{sh}$ is approximately independent of $Q^{2}$
[14]. The leading term of the shadowing
is shown in the last line of Eq.~(\ref{eq:3.1}). For the light
nuclei, it rises with the atomic number $\propto A^{1/3}$:
\beq
1-R_{sh} \propto
{1\over A}\int d^{2}\vec{b} T(b)^{2} \propto {A \over R_{A}^{2}}
\propto A^{1/3} \, .
\label{eq:3.3}
\endeq
For the heavier nuclei, this rise slows down because of the
nuclear attenuation factor in the integrand of (\ref{eq:3.1}).
Evidently, the driving term of the shadowing cross section
is proportional to the forward DD on free nucleons
Eq.~(\ref{eq:2.3}) [4,14,25], and the experimental
observation of nuclear shadowing already is a solid evidence
for significant DD of virtual photons in DIS.

The nuclear profile function for the coherent
DD $\gamma^{*}+A\rightarrow X+A$ equals
\beq
\Gamma_{coh}(\gamma^{*}\rightarrow X,\vec{b})=
\langle X|\{
1-\exp[-{1\over 2}\sigma(r)T(b)]\}
|\gamma^{*}\rangle\, .
\label{eq:3.4}
\endeq
The total cross section of the coherent
DD on a nucleus, integrated over all
final states $X$, equals
\arr
\sigma_{coh}=R_{coh}A\sigma_{D}^{(N)}=
\int d^{2}\vec{b}
\sum_{X}|\Gamma_{coh}(\gamma^{*}\rightarrow X,\vec{b})|^{2}
{}~~~~~~~~~~~~~~~~\nonumber\\ =
\int d^{2}\vec{b} \left< \{
1-\exp[-{1\over 2}\sigma(r)T(b)]\}^{2}\right>
={1\over 4}\int d^{2}\vec{b}T(b)^{2}\left<
\sigma(r)^{2}\exp[-\sigma(r)T(b)]
\right> + ...
\label{eq:3.5}
\endarr
Here the closure was used, and in the last
line of Eq.~(\ref{eq:3.5}) we show the leading term
of the coherent DD cross section, which has a close semblance to the
leading term of the nuclear shadowing in Eq. (\ref{eq:3.1}).

The differential cross section of the incoherent production
$\gamma^{*}+A\rightarrow X+A$ equals
\arr
\left.{d\sigma_{inc}(\gamma^{*}\rightarrow X)\over dt}\right|_{t=0}
=R_{inc}A\left.{d\sigma_{D}^{(N)}\over dt}\right|_{t=0}
{}~~~~~~~~~~~~~~~~~~~~
   \nonumber\\
={1\over 16\pi}
\int d^{2}\vec{b} T(b)\left|\langle X|\sigma(r)
\exp[-{1\over 2}\sigma(r)T(b)]|\gamma^{*}\rangle\right|^{2}
+...
\label{eq:3.6}
\endarr
Here we only
have shown the single incoherent scattering term,
which is sufficient for the practical purposes, the full
multiple-scattering expansion can be found in [22].
Summing over
all produced states $X$ making use of the closure, we can write
\beq
\sigma_{inc}=R_{inc}A\sigma_{D}^{(N)}
=\sigma_{D}^{(N)}
\int d^{2}\vec{b} T(b)
{\langle \sigma(r)^{2}
\exp\left[-\sigma(r)T(b)\right]\rangle
\over
\langle \sigma(r)^{2} \rangle}
\label{eq:3.7}
\endeq

Diffraction excitation of $q\bar{q}$ pairs describes only a
part of nuclear shadowing and of the DD cross section, but
this part is the dominant one at moderate energies and/or large
rapidity gaps. Furthermore,
the $A$-dependence of diffraction excitation
of higher Fock states of the photon will be similar to that
for the $q\bar{q}$ Fock states. In Fig.~1 we present our
predictions for the $A$-dependence of the normalized coherent
$R_{coh}$ and the incoherent $R_{inc}$ diffraction dissociation
of photons into $q\bar{q}$ pairs. We also show the contribution
of excitation of $q\bar{q}$ pairs to nuclear shadowing. The
calculations are based on the dipole cross section of Ref.~[14],
which has already been successfully applied to the quantitative
description of nuclear shadowing in DIS [14,15] and of colour
transparency effects in exclusive leptoproduction
of the $\rho^{0}$ mesons on nuclei [25].

The incoherent and coherent DD have quite a
different $A$-dependence. In the limit of vanishing nuclear
attenuation, when
$\exp[-\sigma(r)T(b)] \approx 1$, we have
$\int d^{2}\vec{b}\,T(b) \exp[-\sigma(r)T(b)]
=A$, and for light nuclei $\sigma_{inc} \approx A\sigma_{D}$ and
 $R_{inc}\approx 1$.
 For heavier nuclei, when the
attenuation factor $\exp[-\sigma(r)T(b)]$
in the integrand of (\ref{eq:3.7}) becomes important,
the relative fraction of
the incoherent production decreases with $A$, i.e.,
$R_{inc} < 1$. As a matter of fact, attenuation is substantial
already for the carbon nucleus, and
for very heavy nuclei
$R_{inc} \ll 1$.
 In the opposite
to that, in the limit of weak attenuation for light nuclei,
the relative fraction of
the coherent diffraction dissociation rises $R_{coh}
\propto A^{1/3}$,
see Eq.~(\ref{eq:3.3}), and the coherent production very rapidly
takes over the incoherent production. In the lightest nuclei
(deuterium,....) the coherent cross section is smaller than the
incoherent one.
For the heavier nuclei
this rise of the coherent cross section slows down because
of the attenuation factor $\exp[-\sigma(r)T(b)]$ in the
integrand of (\ref{eq:3.5}), but persists for the whole range
of nuclei. Furthermore,
for very heavy nuclei, in the black-disc limit,
we predict that the coherent DD makes one half of the total
DIS cross section, see the discussion in Section 6.
A comparison of Eqs.~(\ref{eq:3.5}),(\ref{eq:3.1}) shows
 that for the light nuclei
the coherent production cross section and the nuclear shadowing
cross section are very close to each other:
\beq
R_{coh/Sh}={\sigma_{coh}\over \Delta\sigma_{sh}}=
{R_{coh}\over 1-R_{sh}}\approx 1\, .
\label{eq:3.8}
\endeq
For the heavier nuclei this ratio $R_{coh/Sh}$ slowly decreases
because of stronger nuclear attenuation factor $\exp[-\sigma
(r)T(b)]$ in the coherent production cross section
$\sigma_{coh}$ Eq.~(\ref{eq:3.5})
as compared
to the attenuation factor $\exp[-{1\over 2}\sigma(r)T(b)]$ in
the shadowing cross section $\Delta \sigma_{sh}$ Eq.~(\ref{eq:3.1}).
In Fig.~1 we show the ratio
$R_{D/Sh}$ of the total, coherent plus incoherent,
 DD cross section to the shadowing cross
section:
\beq
R_{D/Sh}={\sigma_{coh}+\sigma_{inc}\over \Delta\sigma_{sh}}=
{R_{coh}+R_{inc} \over 1-R_{sh}}\, .
\label{eq:3.9}
\endeq
 For light nuclei it is slightly above unity for the
sizable contribution of the incoherent DD. The results
shown in Fig.~1
confirm our anticipation of the enhancement
of diffraction production on nuclei with respect to the
free-nucleon target.

%-----------------------------------------------------

%                              Section 4

%--------------------

\section{The energy dependence of diffraction and the
triple-pomeron component of diffraction dissociation}

At high energy, when $\Delta z\gg R_{A}$ and the frozen-size
approximation holds, the diffraction excitation of $q\bar{q}$
pairs gives the energy and/or $x$ independent
  DD cross section. There are two
sources of the energy dependence of DD both on nucleons and
nuclei:

Firstly, at finite energy, in the diffraction excitation
$\gamma^{*}+A\rightarrow X+A$ the target nucleus receives the
longitudinal momentum transfer $\kappa$ Eq.~(\ref{eq:2.4}).
Henceforce, if we ask for the coherent diffraction excitation
when the target nucleus remains in the ground state, then
the production amplitude will be proportional to the so-called
body form factor of the nucleus $G_{A}(\kappa^{2})$, which
for all the practical purposes can be taken equal to the charge
form factor of the nucleus. In the high-energy limit, which
in DIS is the limit of $x\rightarrow 0$, the longitudinal
momentum transfer $\kappa \rightarrow 0$ and $G_{A}(\kappa^{2})
\rightarrow 1$.

At finite energy $\nu$ and/or finite $x$, the effect of this
form factor can be included as follows. Let $d\sigma_{D}^{(N)}
/dM^{2}$ be the mass spectrum of diffraction dissociation on
the free nucleon, in which case the form factor effect can be
neglected for the small size of the nucleon.
For the purposes of the present analysis we can write
\beq
{d\sigma_{coh}\over dM^{2}}
\approx R_{coh}A{d\sigma_{D}^{(N)} \over dM^{2}}G_{A}(2\kappa^{2})
\label{eq:4.1}
\endeq
and
\beq
\sigma_{coh}=R_{coh}A
\int dM^{2}{d\sigma_{D}^{(N)} \over dM^{2}}G_{A}(2\kappa^{2})\, ,
\label{eq:4.2}
\endeq
where the approximation $G_{A}^{2}(\kappa^{2})\approx G_{A}
(2\kappa^{2})$, which is exact for the Gaussian form factor,
was made. Because of this suppression by the nuclear form
factor,
the coherent DD only takes place at a
sufficiently small $x$ such that $\kappa R_{A}\lsim 1$, i.e.,
\beq
x \lsim {1\over R_{A}m_{N}} \sim 0.1A^{-1/3}\, .
\label{eq:4.3}
\endeq
The above simplifying approximation that
nuclear attenuation does not depend
on the mass $M$ of the excited state is viable for the inclusive
DD cross section
(for discussion of possible difference of nuclear attenuation
for the production of specific exclusive final states see [26].
The above form factor suppression is absent in the incoherent
DD on nuclei.

Nuclear shadowing
has its origin in the destructive interference of the
single-scattering (impusle-approximation) amplitude with the
amplitude of the double (and higher order) scattering
$\gamma^{*} \rightarrow  X \rightarrow \gamma^{*}$ [4,14].
The driving term of nuclear shadowing in Eq.~(\ref{eq:3.1})
is proportional to $d\sigma_{D}^{(N)}/dt|_{t=0}$. The principal
finite-energy modification of Eq.~(\ref{eq:3.1}) is as follows:
The intermediate state $X$ acquires the phase $\kappa (z_{2}-
z_{1})$ during its propagation
from the production point $z_{1}$ to the reabsorption
point $z_{2}$. The corresponding contribution to the
double-scattering amplitude enters with the phase factor
$\exp[i\kappa(z_{2}-z_{1})]$ with respect to the impulse
approximation amplitude. After integration over $z_{1,2}$, this
phase factor gives rise to the suppression factor
$G_{A}^{2}(\kappa^{2})\approx
G_{A}(2\kappa^{2})\, $
[4,25,27]. As a result,
nuclear shadowing will be proportional to
\beq
\Delta\sigma_{sh}\propto
\int dM^{2}\left.{d\sigma_{D}^{(N)}\over dtdM^{2}}\right|_{t=0}
G_{A}(2\kappa^{2})\, .
\label{eq:4.4}
\endeq
Consequently, the $x$ dependence of nuclear shadowing
Eq.~(\ref{eq:4.4}) and of the total cross section of the
coherent DD  Eq.~(\ref{eq:4.2}) will be approximately the
same.
One must bear in mind, though, that the magnitude
of the observed
nuclear shadowing is somewhat reduced (by $\sim 5\%$ for
heavy nuclei) because of
 the low-$x$ manifestation of the nuclear EMC
effect [14,15].

Second source of the energy dependence of DD and of nuclear
shadowing is the triple-pomeron component of diffraction
dissociation. Namely, the diffraction excitation of the
$q\bar{q}$ Fock states of the photon gives the mass spectrum
(\ref{eq:2.9}) which converges rapidly at $M^{2}\gsim Q^{2}$.
It is the counterpart of diffraction excitation of resonances
in hadronic interactions. Diffraction excitation of the
$q\bar{q}g$ and higher Fock states generates the so-called
triple pomeron mass spectrum [14,5,6]
\beq
{1\over \sigma_{N}}\left.{d\sigma_{D}^{(N)}\over dtdM^{2}}
\right|_{t=0}
=
{\left[M^{2}\over M^{2}+Q^{2}\right]}^{2} \, .
{A_{3\Pom}\over M^{2}+Q^{2}} \approx
{A_{3\Pom}\over M^{2}+Q^{2}} \, .
\label{eq:4.5}
\endeq
In terms of the structure function of the pomeron, diffraction
excitation of the $q\bar{q}g$ (and higher) Fock states of the
photon corresponds to DIS on the $q\bar{q}$ sea in the pomeron,
the sea being generated from the valense $gg$ component of the
pomeron. The particular mass spectrom (\ref{eq:4.5}), with
the factor $[M^{2}/(M^{2}+Q^{2})]^{2}$ which suppresses the
triple-pomeron contribution at $M^{2}\lsim Q^{2}$, reflects the
large-$\beta$ behaviour of the gluon structure function in
the pomeron $G^{\Pom}(\beta) \propto (1-\beta)$ [5,6].

Although the triple-pomeron coupling $A_{3\Pom}$ is the
dimensionfull constant, it must only weakly depend on
$Q^{2}$ [14,6,28] and can be borrowed from the triple-pomeron
phenomenology of the real photoproduction [29]: $A_{3\Pom}
\approx 0.16$GeV$^{-2}$. This choice of $A_{3\Pom}$ leads to
an excellent quantitative description of the experimental
data on nuclear shadowing [14,15]. For the reference purposes,
in Fig.~2 we present our estimate for the nuclear shadowing
in $\mu Xe$  scattering. The $x$-dependence of $R_{sh}(x)$ at
small $x$ is dominated by the rising contribution to shadowing
from the triple-pomeron component of the mass spectrum.
At $x\gsim 0.01$ the $x$-dependence of shadowing comes
predominatly from
the form factor effects Eq.~(\ref{eq:4.4}).
(Compared to the more detailed analysis
in Ref.~[15], here we neglect corrections for the nuclear EMC
effect, which may reduce nuclear shadowing by $\sim 5\%$. Also,
here we use simple parametrizations (\ref{eq:2.9}), (\ref{eq:4.5})
rather than the direct calculation of the mass spectrum).

In the triple-pomeron regime, the diffraction slope is smaller
than in the resonance and/or $q\bar{q}$ excitation region (for
instance, see [19,29]). Therefore, in our estimates of the DD
cross section we take
$b_{3\Pom}\approx {1\over 2}b_{\pi N} \approx (5-6)$GeV$^{-2}$.

In principle, the triple-pomeron coupling is calculable in
terms of the cross of interaction of the $q\bar{q}g$ Fock
state [6], and such a calculation is in progress. The
same three-parton cross section controls the nuclear attenuation,
which for the $q\bar{q}g$ (and higher) Fock states can be
slightly different from that for the $q\bar{q}$ Fock state.
At very small values of $x$ the triple-pomeron component
of shadowing takes over, but for the moderately small $x$ of the
present muon experiments the dominant contribution to the nuclear
shadowing and to the diffraction dissociation cross section comes
from the $q\bar{q}$ states, see Fig.~2 in which we decompose
nuclear shadowing into the $q\bar{q}$ excitation and triple-pomeron
components. For this reason, for the purposes of the present
analysis, we can make a simplifying assumption of similar nuclear
attenuation of the $q\bar{q}$ and $q\bar{q}g$ states.

Now we are in the position to write down the total mass spectrum
in the diffraction dissociation of photons on a free nucleon:
\arr
F_{N}(y)=
{1\over \sigma_{N}}{d\sigma_{D}^{(N)}\over dy}=
(M^{2}+Q^{2}){d\sigma_{D}^{(N)}\over \sigma_{N} dM^{2}}=
{}~~~~~~~~~~~~~~~~~
\nonumber \\
2r_{D}{Q^{2}M^{2}\over (Q^{2}+M^{2})^{2}}+
{A_{3\Pom}\over b_{3\Pom}}{M^{2} \over Q^{2}+M^{2}}
=2r_{D}{\exp(y)\over [1+\exp(y)]^{2}}+
{A_{3\Pom}\over b_{3\Pom}}{\exp(y)\over 1+\exp(y)}
\, .
\label{eq:4.6}
\endarr
Here the normalization $2r_{D}$ of the $q\bar{q}$ term is written
assuming that $Q^{2}$ is large enough, $Q^{2}\gg M_{min}^{2}$.
For the free-nucleon target, the differential probability of
diffraction excitation of heavy masses $M^{2} \gg Q^{2}$ flattens
at large $y$ at the value
\beq
{d\sigma_{D}^{(N)}\over \sigma_{N}dy}=
{A_{3\Pom}\over b_{3\Pom}} \approx 0.025-0.03\, .
\label{eq:4.7}
\endeq
The triple-pomeron component of the mass-spectrum only becomes
the dominant one at [5]
\beq
M^{2}\gsim (5-7)Q^{2}
\label{eq:4.8}
\endeq
The  $y$-distribution $F_{N}(y)$ for $Q^{2}=1$GeV$^{2}$
and $x=2\cdot 10^{-3}$, relevant to the E665 experiment [13],
is shown in Fig.~3. It is slightly peaked at $y\sim 0.5$
and exhibits the onset of the
triple-pomeron plateau at large $y \gsim 2$, see estimate
(\ref{eq:4.8}).
Excitation of the $q\bar{q}$ pairs dominates at smaller $y$.

We advocate studying the rapidity gap distribution in terms of
the variable $y$, because for the free-nucleon target, and also
for the nuclear target at $x\ll 1$, the $y$ distribution
must be a scaling function of $y$, which depends neither on $Q^{2}$
nor $x$. For the comparison, in Fig.4 we show the predicted
$M^{2}$-distribution for few values of $x$ assuming
$W^{2}= 400$GeV$^{2}$ which is appropriate for the E665 experiment.
In the mass spectrum, this nice $y$-scaling property
is completely obscured.

In the comparison with experimental mass spectra taken
at different values of $x$, one only must bear in mind, that
Eq.~(\ref{eq:4.6}) only holds at the (pseudo)rapidity gap
$\Delta\eta \gsim
2.5-3$, because at smaller
rapidity gaps the non-diffractive mechanisms of the rapidity
gap generation will take over [30]. This puts a restriction on
the excited mass  $M^{2}\lsim 20$GeV$^{2}$ for the $W^{2}\approx
400$GeV$^{2}$ of the E665 experiment. At asymptotically large
$W^{2}$, all the curves must flatten at the same asymptotic value
Eq.~(\ref{eq:4.7}) at large $M^{2}$.
As Eq.~(\ref{eq:2.6}) shows with
increasing $Q^{2}$ and increasing $x$
at the fixed value of $W^{2}$, the less and
less room will be left for the large-$y$
triple-pomeron component and, at
a sufficiently large $Q^{2}$, DD will be dominated by excitation
of the $q\bar{q}$ state. This can also be seen from Fig.~2, in which
we show separately the contribution to nuclear shadowing from
the triple-pomeron component.

%----------------------------------------------------------------------

%                                        Section 5

%------------------------

\section{Experimental signatures of nuclear enhancement of
diffraction dissociation}

%----------------
%              Subsection 5.1

\subsection{The mass spectrum in DD on nuclei and the rapidity-gap
distribution}

The nuclear enhancement/attenuation and the nuclear form factor
effects lead to significant changes in the rapidity gap
distribution:
\arr
F_{A}(y)=
(M^{2}+Q^{2}){d\sigma_{D}^{(A)}\over \sigma_{A} dM^{2}}=
{}~~~~~~~~~~~~~~~~~~~~~~~~~~~\nonumber\\
{R_{inc}+R_{coh}G_{A}(2\kappa^{2})\over R_{sh}(x)}  \cdot
\left\{ 2r_{D}{Q^{2}M^{2}\over (Q^{2}+M^{2})^(x){2}}+
{A_{3\Pom}\over b_{3\Pom}}{M^{2} \over Q^{2}+M^{2}}\right\} =
{}~~~~~~\nonumber\\
{R_{inc}+R_{coh}G_{A}(2m_{N}^{2}x^{2}[1+\exp(y)]^{2})\over R_{sh}(x)}
\cdot
\left\{
2r_{D}{exp(y)\over [1+\exp(y)]^{2}}+
{A_{3\Pom}\over b_{3\Pom}}{\exp(y)\over 1+\exp(y)} \right\}
\label{eq:5.1.1}
\endarr
At small $x \ll 1/R_{A}m_{N}$, the major effect is an enhancement
of large rapidity gaps $\Delta\eta \sim \log({1\over x})$, i.e.,
of excitation of $M^{2}\sim Q^{2}$,  by the factor
\beq
{F_{A}(y)\over F_{N}(y)}=
{R_{coh}+R_{inc}\over R_{sh}(x)} > 1
\label{eq:5.1.2}
\endeq
For the Xe nucleus, used as a target in the Fermilab E665
experiment [13], we predict an enhancement as strong as
$F_{A}(y)/ F_{N}(y)\sim 2.5$, see Fig.~3.
This enhancement comes entirely from the coherent DD, and
is partly due to $R_{sh}(x) <1$.

Because of the form-factor suppression of the coherent DD
cross section, the enhancement decreases with the increase
of $x$, since the minimal longitudinal momentum
transfer (\ref{eq:2.4}) increases with $x$. Similar suppression
takes place with the increase of the mass of the excited
state, i.e., with the increase of $y$ and the decrease of the
rapidity gap $\Delta\eta$. Because the form-factor suppression
is absent in the incoherent DD, at large values of $y$ the rapidity
gap distribution for the nuclear target will be dominated by
the incoherent production and
will flatten at the value which is smaller than for the free
nucleon by the factor $R_{inc}/R_{sh} < 1$:
\beq
{d\sigma_{D}^{(A)}\over \sigma_{A}d\Delta\eta}=
{A_{3\Pom}\over b_{3\Pom}}
{R_{inc}\over R_{sh}(x)}\, .
\label{eq:5.1.3}
\endeq
In Fig. 3  we show the $y$-distribition for the $\mu Xe$ interactions
for few values of $x$ assuming $W^{2}=400$GeV$^{2}$.
As a matter of fact, in the kinematical range of the E665
experiment, the form factor effects are quite significant and
there is no  room for the flat $y$-distribution (\ref{eq:5.1.3})
at large $y$.
Consider, for instance, $W^{2}=400$GeV$^{2}$
and $Q^{2}=2$GeV$^{2}$, i.e., $x=5\cdot 10^{-3}$. For the
$Xe$ nucleus the charge radius $R_{ch}\approx 4.5$f [24]. The
suppression by the form factor becomes significant at
$R_{ch}m_{N}x[1+\exp(y)] \gsim 1$, i.e., at
$M^{2}/Q^{2} \gsim 9$ and $y\gsim 2$. This corresponds
to the onset of the triple-pomeron region, but still larger
values of $y$ are needed to reach the dominance of the incoherent
DD. However, at this value
of $x$, the requirement of $\Delta \eta \gsim 3$ imposes the
upper bound $y \lsim 2.3$, see Eq.~(\ref{eq:2.9}). Even at
$x=2\cdot 10^{-3}$ the triple-pomeron plateau is still elusive.
At still larger values of $x$, the form factor effects become
important at smaller values of $y$. This leads to the
non-scaling $F_{A}(y)$  for DD on nuclei, whereas in DD on
the free nucleons $F_{N}(y)$ is predicted to not depend on
$x$ and $Q^{2}$.
The nuclear enhancement of large rapidity gaps, of small $y$
and  small
$M^{2}$, and nuclear suppression of smaller rapidity gaps,
i.e., of large
$y$, is a very specific prediction, which can easily be tested
experimentally. As we stated above, because of finite energy
only the former prediction
can be tested in the energy region of the E665 experiment.

In Fig.~5  we present the fraction of diffraction dissociation
\beq
W_{D}(y^{*})=\int^{y^{*}}dy F(y)
\label{eq:5.1.4}
\endeq
as a function of $y^{*}$ at different values of $x$.
In Fig.~6 we present our prediction
for the total, coherent plus incoherent, rate of DD
$w_{D}(x)$ as a function of $x$ for the free-nucleon
and $Xe$ targets for the rapidity-gap cutoff
$\Delta\eta\gsim 3$. Bear in mind that the upper bound on
$y^{*}$ changes with $x$.
At small $x$, the fraction of DD and/or LRG events on
a nucleus is significantly higher than for the free-nucleon target.
At large $x$, fractions of DD in $\mu Xe$ and $\mu N$ interactions
converge. Because of the nuclear form factor effects, nuclear
coherent DD vanishes at $x\gsim 0.1$, and here
a probability of LRG events in the
$\mu Xe$ case even
will be smaller than in the $\mu N$ case.
For the nucleon target, apart from the $x$-dependence of the
width of
the allowed region of $y$,
the probability $W_{D}(y)$ only weakly
depends on $x$ and $Q^{2}$.
For the nuclear target there is a slight violation of this
scaling, related to the scaling violation by the nuclear
form factor effects in $F_{A}(y)$.

Notice, that whereas the diffraction dissociation is necessarily
accompanied by the formation of a large (pseudo)rapidity gap, the
reverse is not necessarily true. Specifically, even in
the non-diffractive, valence dominance region of DIS, the
virtual photon can fragment into the hadronic system of very
small mass $M^{2} \ll Q^{2}$, which will be separated from the
recoil nucleon by large (pseudo)rapidity gap $\Delta\eta =
\log(W^{2}/M^{2})$. Such a non-diffractive rapidity-gap events must
have an origin in the multiplicity fluctuations [30] and/or the
secondary reggeon exchange across the rapidity gap, and must have
the differential rapidity gap distribution which decreases
exponentially at large rapidity gap,
\beq
F_{ND}(y) \propto \exp(-\Delta\eta) \propto x\cdot \exp(y)
\label{eq:5.1.5}
\endeq
compared to the flat or even rising probability
of large diffractive gaps $\Delta \eta \sim \log({1\over x})$,
i.e., of $y\sim 0$. Indeed, we predict quite a striking rise of
$F_{A}(y)$ towards $y\rightarrow 0$, see Fig.~3.

The preliminary results from the E665 experiment on $\mu Xe$
interactions [31] confirm the above predictions. The
E665 defines the large-rapidity gap (LRG) events subject to the
psedurapidity gap $\Delta \eta > 2$, and separates the
total statistics into the shadowing $x \leq 0.02$ and non-shadowing
$x\geq 0.02$ samples. Their free-nucleon sample comes from
the $\mu D$ interactions. The E665 lower bound for the
fraction of DD
is $0.12\pm 0.02$ for the $\mu D$ and $0.18\pm 0.03$ for the
$\mu Xe$ interactions in the shadowing region [31], which
is consistent with our prediction of nuclear enhancement of DD.

%----------------------------

%                       Subsection  5.2

\subsection{DD and grey tracks}

The nuclear multiproduction events are conveniently classified
according to the multiplicity $n_{g}$ of the so-called grey
tracks, which are predominantly recoil protons with the momentum
$(150-200)\lsim p\lsim 600\,{\rm MeV/c}$. Here the lower cutoff
is usually so chosen as to exclude the nucleus evaporation products.
The multiplicity of grey tracks $n_{g}$ measures the mutiplicity
of inelastic intranuclear interactions. Still another similar
observable is the total charge $Q_{T}$ of secondary hadrons. In
interaction with the free-nucleon (deuteron) target $\langle Q_{T}
\rangle={1\over 2}$. The $Q_{T} \geq 2$ in multiproduction
on a nucleus is an unambiguous signature of intranuclear cascading.

Evidently, the coherent DD on a nucleus entirely falls
into the $n_{g}=0$ and/or $Q_{T}=0$ cathegory. In the incoherent
diffraction dissociation the longitudinal recoil $\kappa$ is smaller
than the lower cutoff for the grey tracks. The transverse
recoil momentum
\beq
p_{\perp} \sim b_{D}^{-1/2}, b_{3\Pom}^{-1/2} \sim 300-400\,{\rm MeV/c}
\label{eq:5.2.1}
\endeq
can be sufficiently large to partly contribute to the $n_{g}\geq 1$,
$Q_{T} \geq 1$ cathegory. However, as we have seen above, the
incoherent DD only makes a small fraction of DD on a nucleus,
see Fig.~1. Consequently, the above predicted nuclear
emhancement of DD leads to a prediction of the
decrease of $\langle n_{g}\rangle $, $\langle Q_{T}\rangle$ with
the decrease of $x$ from the nonshadowing (NS)
region of $x\gsim
1/R_{A}m_{N}$ to the shadowing (Sh)
region of $x \ll 1/R_{A}m_{N}$.
To a crude approximation, we may assume that the multiplicity of
grey tracks and/or the total observed charge in the non-diffractive
interactions do not depend on $x$ , which leads to the estimate
\arr
\langle n_{g} \rangle  _{Sh} \approx
[1-W_{D}(x)]\langle n_{g} \rangle  _{NS} \,, \nonumber \\
\langle Q_{T} \rangle  _{Sh} \approx
[1-W_{D}(x)]\langle Q_{T} \rangle _{NS} \,.
\label{eq:5.2.2}
\endarr
Our estimate for the suppression factor $1-W_{D}(x)$ is shown
in Fig.~7.
Notice, that in the hadron-nucleus interaction $\langle n_{g}\rangle$
rises with $\bar{\nu}$, which by the simple-minded extrapolation
would have suggested the rise of $\langle n_{g}\rangle$ from the
non-shadowing region of large $x$ to the shadowing region of small
$x$ [1,2,8-10].
As a matter of fact, this study was primarily motivated
by the preliminary evidence for such a
$x$-dependence of $\langle n_{g} \rangle$ in the E665 data on
$\mu Xe$ interaction [31]. The E665 data show an $\approx 30\%$
reduction of $\langle n_{g} \rangle$ and $Q_{T}$ from $x\gsim 0.1$
to $x\sim 0.001$,  which is in very good agreement with
the estimate (\ref{eq:5.2.2}) shown in Fig.~7.
One would expect [8-10] stronger intranuclear cascading and
certain enhancement
of $\langle n_{g}\rangle$ going from the non-shadowing to
the shadowing region of $x$, so that Eq.~(\ref{eq:5.2.2})
gives rather the lower bound for the suppression factor.

The fraction of diffractive production $W_{D}(x)$ is a scaling
function of $x$, and we advocate binning the experimental data
{\sl vs.} $x$ rather than {\sl vs.} $Q^{2}$ and/or $W^{2}$.
 Small values of $x$ are only accessible at high energy $\nu$,
and we predict a decrease of $\langle n_{B}\rangle$, $\langle n_{g}
\rangle$ and $\langle Q_{T}\rangle $ with increasing energy.

%--------------------

%                   Subsection 5.3

\subsection{DD and the (pseudo)rapidity spectrum of secondary
particles}

In the non-difractive (ND) inelastic interaction, secondary
particles populate the whole rapidity span. By the formation-time
considerations, the forward particle production in the ND events
must be target independent [1,8-10].
In the DD events
the secondary paricles populate only the photon hemisphere,
with the vanishing activity in the target region for the
coherent DD, and with some signal from the recoil protons
in the incoherent DD.
Now we comment in more detail on the
signal of nuclear enhancement of DD in the (pseudo)rapidity
spectra. The small effects of DD on the very forward hadroproduction
on nuclei were discussed earlier in [18].

A convenient quantity is
the normalized (pseudo)rapidity $\eta$ distribution
of secondary particles in the photon-proton center-of-mass system
$$
R(\eta)=\left({dn^{(A)} \over d\eta }\right)
\left({dn^{(N)} \over d\eta }\right)^{-1}  \, ,
$$
and let $\eta > 0$ be the photon hemisphere. In the generic
particle-nucleus interaction $R(\eta)$ rises towards the target
fragmentation region because of the nuclear cascading [1,2,8-12].
In the
hadron-nucleus interactions, because of simultaneous interactions of
few constituent quarks of the hadron, in the central region $R(\eta)
> 1$, and in the prejectile fragmentation  region $R(\eta) < 1$ (for
the review see [1]). In the leptoproduction on nuclei in the
photon fragmentation region $R(y^{*}) \approx 1$ with some
evidence for nuclear depletion in the maximal-$y^{*}$ bin [12,13].

Evidently, the nuclear enhancement of DD leads to a higher
particle density in the forward hemisphere in the multiproduction
on nuclei compared to the free nucleon (deuterium) target.
However, this simple prediction is not easy to test. Firstly,
the total rapidity span depends on $W^{2}$. Because
the rapidity spectrum $dn^{(N)}/d\eta$ is a steep function of
$\eta$, small mismatch in $W^{2}$ may lead to large spurious
effects in $R(\eta)$. Secondly, the diffraction dissociation
products are smeared over broad rapidity range $\propto \log(M^{2})$
and such a smearing is even broader in the pseudorapidity variable.
The effect of the smearing tends to diminish the departure of
$R(\eta)$ from unity [1]. Very crude estimate for
forward production is
\beq
R_{A/D}(\eta) - 1  \approx W_{D}^{(A)}(x)-W_{D}^{(N)}(x)
\label{eq:5.3.1}
\endeq
Notice, that because of the nuclear suppression of DD in the
nonshadowing region of large $x$, in the non-shadowing region
we expect $R_{A/D}(\eta) <1$ for the very forward particles.

One may also compare the nuclear interactions in the non-shadowing
and shadowing region, where one would expect
\beq
R_{sh/NS}(\eta) - 1  \approx W_{D}^{(A)}(x)
\label{eq:5.3.2}
\endeq
Here one compares the spectra at different values of $Q^{2}$, and
the effect can be masked by weak $Q^{2}$ dependence of the
rapidity spectra. The effect can somewhat be enhanced, if one
compares the forward pseudorapidity spectra in the $n_{g}=0$ events
in the shadowing and non-shadowing regions on the same nucleus.
The sample of the $n_{g}=0$ events will evidently be enriched
by DD. If $P_{NS}(0)$ is the probability of having $n_{g}=0$
for the nonshadowing region of large $x$, then in the shadowing
region we expect
\beq
P_{sh}(0)\approx P_{NS}(0)+w_{D}(x) \, .
\label{eq:5.3.3}
\endeq
Whith the above reservations about possible $Q^{2}$ dependence
of the spectra, we expect
\beq
R_{sh/NS}(n_{g}=0, \eta) \approx 1 + {w_{D}(x) \over P_{NS}(0)}
\label{eq:5.3.4}
\endeq
Similar effect is expected, if one compares the forward production
in the $n_{g}=0$ and the $n_{g}\geq 1$ samples.

In all the above cases, this enhancement can, perhaps, best be
seen by a comparison of average multiplicities in the forward
hemisphere $\langle n_{F} \rangle$. For instance, in view of
Eq.~(\ref{eq:5.3.3})  we predict that $\langle n_{F}\rangle$ in the
$n_{g}=0$ sample of $\mu Xe$ interactions must be larger than for
the free-nucleon target. We expect a similar enhancement of the
forward multiplicity in the shadowing region of small $x$ compared
to the non-shadowing region of large $x$, although such a comparison
is somewhat indirect because of different values of $Q^{2}$ in the
two regions and possible slight dependence of mean multiplicty on
$Q^{2}$. Similar enrichement by diffraction dissociation must
hold also for the forward production in the $Q_{T}=0$ sample
of nuclear interactions.

Consider now the multiplicities in the target hemisphere. Here
the intranuclear cascading leads to $R_(\eta)> 1$ in the global
rapidity spectra. Because DD does not contribute to the target
hemisphere, the rising fraction $w_{D}(x)$ of DD must be
associated with the decrease of the mean multiplicity
$\langle n_{B} \rangle$ of secondary partilces produced in the
nucleus hemisphere, cf. Eq.~(\ref{eq:5.2.2}):
\arr
{\langle n_{B} \rangle  _{Sh} \over \langle n_{B} \rangle  _{NS} }
\approx 1-w_{D}(x)
\label{eq:5.3.5}
\endarr
This estimate is in very good agreement with the preliminary
data from the E665 experiment [31], which found $\approx 30\%$
depletion of $\langle n_{B} \rangle$ from $x\gsim 0.1$ to
$x\sim 0.001$, and the observed depletion is the same as
for $\langle n_{g}\rangle$ and $Q_{T}$.

%-----------------------------------------------------

%                                      Section 6

\section{What makes the hadroproduction and leptoproduction on nuclei
different?}

Diffraction dissociation and LRG interactions exist in the hadronic
interactions too, but their effect on the multiproduction on nuclei
is marginal. The principal distinction between the leptoproduction
and hadroproduction is strong absorption via elastic rescatterings
in the latter case. Furthermore, the diffraction dissociation of
hadrons makes only a small fraction of diffractive scattering,
which is dominated by elastic scattering.

Let us make this argument more explicit. The formalism of
Section 2 is fully applicable to diffractive scattering of
hadrons. The total cross section of $hN$ scattering can be
written as  $\sigma_{tot}(hN) =\langle \hat{\sigma} \rangle_{h}$,
where the subscript $h$ denotes the matrix element (\ref{eq:2.2})
over the wave function of the hadron $h$ and
$\hat{\sigma}$ stands for the generic cross section
operator, $\hat{\sigma}=\sigma(r)$ in the example considered
in Section 2. The diffrential cross section of the forward
elastic scattering can then be written as (we suppress the
subscript $h$)
\beq
\left.{d\sigma_{el} \over dt}\right|_{t=0} = {1\over 16\pi}
\langle \hat{\sigma} \rangle ^{2} \, .
\label{eq:6.1}
\endeq
If all eigenvalues of the diffraction matrix were identical,
then the diffractively scattered wave would have differed from
the incoming hadronic wave only by the overal phase/attenuation
factor, and there would not have been any diffraction dissociation
(for the review see [1]). The counterpart of Eq.~(\ref{eq:2.3})
for hadrons is [1,22]
\beq
\left.{d\sigma_{el} \over dt}\right|_{t=0} = {1\over 16\pi}
\left(\langle \hat{\sigma}^{2} \rangle -
\langle \hat{\sigma} \rangle ^{2}\right)
\label{eq:6.2}
\endeq
and DD measures the dispersion of eigenvalues of the diffraction
scattering operator. For the proton-proton scattering,
the detailed analysis of the DD data gives  [32]
\beq
{\langle \hat{\sigma}^{2} \rangle -
\langle \hat{\sigma} \rangle ^{2}\over
\langle \hat{\sigma} \rangle ^{2}} \sim 0.3 \, .
\label{eq:6.3}
\endeq
At high energy, elastic scattering makes $\sim 20\%$ of the
total $pp$ cross section, and single-arm DD makes only $\sim
6 \%$ of the total cross section. In DIS at small $x$ we
found much larger fraction of DD, $\sim 15\%$ in the $\mu N$
scattering and $\sim (35-40)\%$ in the $\mu Xe$ scattering
at $x=0.002$. Is this reasonable?

In the case of photons the term
$\langle \hat{\sigma}\rangle^{2}$ is negligibly small, as it
contains the extra power of the fine structure constant
 $\alpha_{em}=1/137$. Then, the
comparison of Eq.~(\ref{eq:2.3}) with
Eqs.~(\ref{eq:6.2},\ref{eq:6.3})
shows that the strength of DD in the photoabsorption
corresponds to the combined strength of elastic scattering and
the beam DD in the hadronic scattering. This explains why
we find such a strong DD for photons. Furthermore, consider
the total photoabsorption cross section Eq.~(\ref{eq:3.1})
and the coherent DD cross section Eq.~(\ref{eq:3.5}) in the
limit of very heavy nucleus, when the absorption becomes
strong. In this limit one will find that the coherent DD on
a nucleus makes $~{1\over 2}$ of the total cross section,
\beq
\sigma_{coh} \sim {1\over 2}\sigma_{A} \, ,
\label{eq:6.4}
\endeq
which
precisely corresponds to the black disc limit when the elastic
cross section equals half of the total cross section.

The above rise of the coherent DD cross section to
$\sim {1\over 2}$
of the total DIS cross section
must be contrasted with very small cross of DD of hadrons on
nuclei, which simply vanishes in the black-disc limit.
The principal point is the following one. The above
estimate (\ref{eq:6.3}) shows that dispersion of the cross
section for Fock states of the hadron is not large.
Then, in the
hadron-nucleus scattering, the overall nuclear attenuation
will be dominated by the average value of the cross section,
i.e., by the free-nucleon cross section. In other words, it
will be dominated by
elastic rescatterings
 of the projectile hadron, and the Glauber formula
for the nuclear total cross section [3]
\beq
\sigma_{hA}=2\int d^{2}\vec{b}\{
1-\exp[-{1\over 2}\sigma_{hN}T(b)]\}
\label{eq:6.5}
\endeq
gives an excellent description of the nuclear shadowing, which in
the $nA$ scattering on heavy nuclei reduces the total cross section
by more than the factor 2 (for analysis of nuclear shadowing in
hadron-nucleus scattering see [7]). The diffraction dissociation
effects are present in this case too and contribute to the nuclear
shadowing. The corresponding correction to the nuclear shadowing,
usually referred to as Gribov's inelastic shadwoing, can be
evaluated as [1,4,7,24]
\arr
\Delta\sigma_{sh}=
4\pi\left.{d\sigma_{D}^{(N)}\over dt}\right|_{t=0}
\int d^{2}\vec{b}T(b)^{2}
\exp[-{1\over 2}\sigma_{hN}T(b)]
+ ...
\, .
\label{eq:6.6}
\endarr
and only makes 2-4\% correction to the impulse approximation cross
section and 5-7\% correction to the total $nA$ cross section [5,22].
The cross section of the diffraction dissociation on nuclei will be
even smaller because of the stronger nuclear attenuation factor:
\arr
\sigma_{D}=
4\pi\left.{d\sigma_{D}^{(N)}\over dt}\right|_{t=0}
\int d^{2}\vec{b}T(b)^{2}
\exp[-\sigma_{hN}T(b)]
+ ...
\, .
\label{eq:6.7}
\endarr
For this reason, LRG events only contribute
a negligible fraction to
the hadronic multiproduction on nuclei, and
their effect is only
noticable for very fast particle production [18].
Significant nuclear enhancement of the mean
multiplicity in the hadroproduction on
nuclei comes about equally from
reinteractions of the projectile and from
the cascading effects [33,34].

%-------------------------------------------------------------

%                                     Section 7

\section{Conclusions}

Our principal finding is a strong nuclear enhancement of the
diffraction dissociation of (virtual) photons in deep inelastic
leptoproduction
off nuclei. This
enhancement of diffraction dissociation amounts to enhancement of
the large-rapidity gap and/or small multiplicity events in the
nuclear shadowing region, in which photons are expected to have
stronger hadronic properties. The nuclear enhancement of the
coherent diffraction dissociation also amounts to suppression of
the target-nucleus fragmentation. Henceforth, a somewhat paradoxial
conclusion that the
more hadronlike photons produce less particles on
nuclei. The principal difference from the hadron-nucleus interactions
is that in the $hA$ case the nuclear shadowing is dominated by the
elastic rescatterings of the projectile hadron, and that the
diffraction dissociation effects, alias the inelastic shadowing,
only make a very small correction to the nuclear cross section.
Diffraction dissociation of hadrons vanishes for the black nuclei.
In contrast to that, in deep inelastic scattering on the completely
absorbibg, black, nuclei the coherent diffraction dissociation must
make $\sim {1\over 2}$ of the total nuclear DIS
cross section.
\vspace{1cm}\\

{\bf Acknowledgements:} One of the authors (N.N.N.) is grateful to
Ivo Derado and Wolfgang Wittek for invitation to visit MPI, Munich
in December 1993, when this work started being inspired by
discussions on the preliminary
results of the E665 experiment [31].
This work was partially
supported by the INTAS grant 93-239. The work of V.R.Z. was
partially supported by the G.Soros International Science Foundation
grant  N MT5000.
\pagebreak\\

{\bf \Large Figure captions:}
\begin{itemize}

\item[Fig.1]
 - The $A$-dependence of the contribution from interaction of the
$q\bar{q}$ Fock state of the photon to the
  DD related quantities: nuclear shadowing
 $R_{sh}$ is shown by the dotted curve; the nuclear enhancement
of the coherent DD cross section $R_{coh}$ and suppression of the
incoherent DD cross section $R_{inc}$ are shown by the dashed and
dot-dashed curve; the solid curve is for the ratio of the total
DD cross section to the shadowing cross section. The definitions
are given in the text
Eqs.~(\ref{eq:3.1},\ref{eq:3.5},\ref{eq:3.6},\ref{eq:3.9}).

\item[Fig.2]
- Nuclear shadowing in $\mu Xe$ scattering as a function of $x$.
The dashed line shows the contribution to shadowing from the
$q\bar{q}$ Fock state of the photon, the solid line includes also
the triple-pomeron component from shadowing of higher Fock states.

\item[Fig.3]
- The diffraction dissociation mass spectrum
Eqs.~(\ref{eq:4.6},\ref{eq:5.1.1})
as a function of
the scaling variable $y=\log(M^{2}/Q^{2})$
The solid curve is for $\mu N$ interaction at $x=0.002$
the dashed, dotted and dash-dotted curves are for $\mu Xe$
interaction at $W^{2}=400$GeV$^{2}$ and
different values of $x$.

\item[Fig.4]
- The diffraction dissociation mass spectrum in $\mu N$
interaction at $W^{2}=400$GeV$^{2}$ and
different values of $x$ is shown as a function of $M^{2}$.

\item[Fig.5]
- The fraction of diffraction dissociation Eq.~(\ref{eq:5.1.4})
integrated over excited masses $M^{2}\leq Q^{2}\exp(y^{*})$
at $W^{2}=400$GeV$^{2}$ and different values of $x$.
The end points of curves correspond to the rapidity gap
$\Delta\eta=3$.

\item[Fig.6]
- The $x$-dependence of the
fraction of diffraction dissociation
integrated over rapidity gaps $\Delta\eta \geq 3$
at $W^{2}=400$GeV$^{2}$.

\item[Fig.7]
- The estimate of suppression of grey particle multiplicity
because of nuclear enhancement of DD.
\end{itemize}

\begin{thebibliography}{99}

\bibitem{1}
N.N.~Nikolaev, {\sl Sov. Phys. Uspekhi} {\bf 24(7)} (1981) 531;
{\sl Sov. J. Part. Nucl.} {\bf 12} (1981) 63.

\bibitem{2}
N.N.~Nikolaev and V.I.~Zakharov. {\it Phys.Lett.} {\bf B55} (1975) 397;
V.I.~Zakharov and N.N.~Nikolaev. {\it Sov.J.Nucl.Phys.} {\bf 21} (1975)
227;

\bibitem{3}
R.J.Glauber, in: {\sl Lectures in Theoretical Physics}, v.1.,
eds. W.Brittain and L.G.Dunham. Interscience Publ., N.Y., 1959;
R.J.Glauber and G.Matthiae, {\sl Nucl. Phys.} {\bf B21} (1970) 135.

\bibitem{4}
V.N.Gribov, {\sl Sov. Phys. JETP} {\bf 29} (1969) 483.

\bibitem{5}
N.N.Nikolaev and B.G.Zakharov, {\sl Z. Phys.}, {\bf C49} (1991)
 607; {\bf C53} (1991) 331.

%-----------------------------   5  --------------

\bibitem{6}
N.N.Nikolaev and B.G.Zakharov, {\sl JETP}, {\bf 78(5)} (1994).

\bibitem{7}
N.N.Nikolaev, {\sl Z. Phys.} {\bf C32} (1986) 537.

\bibitem{8}
N.N.~Nikolaev, {\sl JETP Letters} {\bf 22} (1975) 419.
G.V.Davidenko and N.N.Nikolaev, {\sl Sov. J. Nucl. Phys.}
{\bf 24} (1976) 402.

\bibitem{9}
N.N.Nikolaev, {\sl Z. Phys.} {\bf C5} (1980) 291.

\bibitem{10}
G.V.Davidenko and N.N.Nikolaev, {\sl Nucl. Phys.} {\bf B135} (1978) 333.

%----------------------------     10   --------------

\bibitem{11}
N.N.Nikolaev and V.R.Zoller, {\sl Nucl. Phys.} {\bf B147} (1979) 336.

\bibitem{12}
N.N.Nikolaev and A.Tenner, {\sl Nuovo Cim.} {\bf A105} (1992) 1001.

\bibitem{13}
E665 Collaboration: M.R.Adams et al., {\sl Z. Phys.} {\bf C61}
(1994) 179;

\bibitem{14}
N.N.~Nikolaev and B.G.~Zakharov. {\it Z.Phys.} {\bf C49} (1991) 607.

\bibitem{15}
V.~Barone, M.~Genovese, N.N.~Nikolaev, E.~Predazzi and B.G.~Zakharov,
{\sl Z. Phys.} {\bf C58} (1993) 541.

%-------------------------------     5  ---------------

\bibitem{16}
M.R.Adams et al., {\sl Phys. Rev. Lett.} {\bf 68} (1992) 3266;
{\sl Phys. Lett.} {\bf B287} (1992) 375.


\bibitem{17}
ZEUS Collaboration: M.Derrick et al., {\sl Phys. Lett.} {\bf B315}
(1993) 481; {\bf DESY 94-063} (1994).

\bibitem{18}
V.R.Zoller, {\sl Z. Phys.} {\bf C44} (1989) 645.

\bibitem{19}
N.N.Zotov and V.A.Zarev, {\sl Sov. Phys. Uspekhi.} {\bf 51} (1988) 119.

\bibitem{20}
N.N.Nikolaev and V.R.Zoller, {\sl Z. Phys.} {\bf C56}
(1992) 623.

%---------------------------  20  ------------

\bibitem{21}
V.~Barone, M.~Genovese, N.N.~Nikolaev, E.~Predazzi and B.G.~Zakharov,
{\sl Phys. Lett.} {\bf B326} (1994) 161.

\bibitem{22}
N.N.Nikolaev, {\sl Sov. Phys. JETP} {\bf 54} (1981) 434.


\bibitem{23}
Al.B.Zamolodchikov, L.I.Lapidus and B.Z.Kopeliovich, {\sl JETP Letters}
{\bf 33} (1981) 612.

\bibitem{24}
H. de Vries, C.E. de Jager and C. de Vries, {\sl Atomic Data
and Nuclear Data Tables} {\bf 36} (1987) 495.

\bibitem{25}
V.A.Karmanov and L.A.Kondratyuk, {\sl JETP Letters} {\bf 18} (1973) 266.

%-----------------------------    25   ----------

\bibitem{26}
B.Z.Kopeliovich and B.G.Zakharov, {\sl Phys. Rev.} {\bf D44}
(1991) 3466;
B.Z.Kopeliovich, J.Nemchik, N.N.Nikolaev and B.G.Zakharov,
{\sl Phys. Lett.} {\bf B324} (1994) 469; {\bf B309} (1993) 179.

\bibitem{27}
N.N.NIkolaev, A.Szczurek, J.Speth, J.Wambach, B.G.Zakharov and V.R.Zoller,
{\sl Nucl. Phys.} {\bf A567} (1994) 781.

\bibitem{28}
N.N.~Nikolaev. {\it Oxford Univ. preprint}
{\bf 58/84} (1984); Also in: {\it Multiquark Interactions and
Quantum Chromodynamics.} Proc. VII Intern. Seminar on Problems
of High Energy Physics, 19-26 June 1984, Dubna, USSR.

\bibitem{29}
T.J.~Chapin et al. {\it Phys.Rev} {\bf D31} (1985) 17.

\bibitem{30}
N.N.Nikolaev and V.R.Zoller, {\sl Sov. J. Nucl. Phys.} {\bf 36}
(1982) 897.

%-------------------------------  30  -------

\bibitem{31}
W.Wittek, Nuclear shadowing and diffraction dissociation in muon-xenon
interactions at 490 GeV. {\sl XXIX-th Rencontre de Moriond, "QCD
and high-energy hadronic interactions",} Meribel, France, March 19-26,
1994;
E665 Collaboration: M.R.Adams et al., Dependence of hadron production
on the number of grey tracks in $\mu Xe$ interactions at 490 GeV,
{\sl paper in preparation}.

\bibitem{32}
L.G.Dakhno and N.N.Nikolaev, {\sl Nucl. Phys,} {\bf A434  }
(1986) 653.

\bibitem{33}
B.B.Levchenko and N.N.Nikolaev, {\sl Sov. J. Nucl. Phys.} {\bf 42}
(1985) 1255, and references therein;

\bibitem{34}
N.N.Nikolaev, {\sl Z. Phys.} {\bf C44} (1989) 645.
\pagebreak\\
\end{thebibliography}
\end{document}